\documentclass[seceqn]{elsart}

\usepackage{graphicx}

\usepackage{amssymb}
\usepackage{mathrsfs}

\usepackage{amsmath}
\usepackage{type1cm}
\usepackage{bbm}

\DeclareMathOperator{\tr}{tr}

\newcommand{\ring}{\mathaccent"7017 }

\begin{document}

\begin{frontmatter}



\title{SUSY WT identity in a lattice formulation of 2D $\mathcal{N}=(2,2)$ SYM}


\author{Daisuke Kadoh},
\ead{kadoh@riken.jp}
\author{Hiroshi Suzuki}
\ead{hsuzuki@riken.jp}
\address{Theoretical Physics Laboratory, RIKEN, Wako 2-1, Saitama 351-0198,
Japan}

\begin{abstract}
We address some issues relating to a supersymmetric (SUSY) Ward-Takahashi
(WT) identity in Sugino's lattice formulation of two-dimensional (2D)
$\mathcal{N}=(2,2)$ $SU(k)$ supersymmetric Yang-Mills theory (SYM). A
perturbative argument shows that the SUSY WT identity in the continuum theory
is reproduced in the continuum limit without any operator
renormalization/mixing and tuning of lattice parameters. As application of the
lattice SUSY WT identity, we show that a prescription for the hamiltonian
density in this lattice formulation, proposed by Kanamori, Sugino and~Suzuki,
is justified also from a perspective of an operator algebra among
correctly-normalized supercurrents. We explicitly confirm the SUSY WT identity
in the continuum limit to the first nontrivial order in a semi-perturbative
expansion.
\end{abstract}

\end{frontmatter}

\section{Introduction and the summary}
\label{sec:1}
In the present note, we derive an identity in Sugino's lattice formulation of
two-dimensional (2D) $\mathcal{N}=(2,2)$ supersymmetric $SU(k)$ Yang-Mills
theory (SYM)~\cite{Sugino:2003yb,Sugino:2004qd} that would become the
supersymmetric (SUSY) Ward-Takahashi (WT) identity in the continuum limit.
(We term this identity a lattice SUSY WT identity for brevity.) On the basis of
formal perturbation theory, we then address the renormalization and mixing
of composite operators appearing in the identity. Our consideration is quite
parallel in the spirit to the standard analysis of the chiral symmetry on the
lattice~\cite{Bochicchio:1985xa}. Compared with the four-dimensional cousin,
four-dimensional $\mathcal{N}=1$
SYM~\cite{Curci:1986sm,Donini:1997hh,Taniguchi:1999fc,Farchioni:2001yr,%
Farchioni:2001wx,Montvay:2001aj}, the situation in 2D $\mathcal{N}=(2,2)$ SYM
is much simpler or almost trivial, because this 2D model is
super-renormalizable. We can in fact argue that, in the continuum limit, the
lattice SUSY WT identity reproduces the SUSY WT identity in the continuum
target theory without any operator renormalization/mixing and tuning of
parameters. This conclusion is consistent with the expected SUSY restoration
without fine-tuning in the effective action of elementary fields, which has
been discussed within perturbation theory~\cite{Sugino:2003yb}. That
consideration on the SUSY restoration in~Ref.~\cite{Sugino:2003yb} implies the
restoration of the SUSY WT identity in the continuum limit was claimed
in~Ref.~\cite{Kanamori:2008bk} only intuitively. The present analysis remedies
this gap.

As an interesting application of the lattice SUSY WT identity, we show that a
prescription for the hamiltonian density in this lattice formulation, advocated
in~Refs.~\cite{Kanamori:2007ye,Kanamori:2007yx} in the context of the
spontaneous SUSY breaking, can be justified also from a perspective of a
``current'' algebra among supercurrents and the hamiltonian density (our
argumentation is analogous to that for the order parameter of the spontaneous
chiral symmetry breaking in~Ref.~\cite{Bochicchio:1985xa}). For other numerical
application of the present lattice formulation, see
Refs.~\cite{Kanamori:2008yy,Kanamori:2009dk,Hanada:2009hq}.

Our argument to this stage is standard but somewhat formal. To partially
substantiate our formal argument, we carry out a one-loop calculation that
confirms the SUSY WT identity in the first nontrivial order of a
semi-perturbative expansion~\cite{Onogi:2005cz} which is justified for small
volume lattices.

The present lattice formulation is based on the A~model topological twist of
2D $\mathcal{N}=(2,2)$ theories~\cite{Witten:1993yc}. For 2D
$\mathcal{N}=(2,2)$ $U(k)$ SYM, there exists another type of lattice
formulation, proposed initially by~Ref.~\cite{Cohen:2003xe} and independently
in~Ref.~\cite{D'Adda:2005zk}, that can be understood in terms of the B-model
topological twist. For this another type of lattice formulation and related
works, see Ref.~\cite{Catterall:2009it} for a recent review,
Ref.~\cite{Kadoh:2009yf} and references cited therein.

\section{Lattice SUSY WT identity}
\label{sec:2}
A most salient feature of the lattice formulation
of~Refs.~\cite{Sugino:2003yb,Sugino:2004qd} is that it is exactly invariant
under a fermionic symmetry~$Q$, defined by\footnote{In this note, we adopt the
convention that all lattice variables are dimensionless. The mass dimension of
fields in the continuum theory is provided by multiplying appropriate powers
of the lattice spacing.}
\begin{align}
   &QU_\mu(x)=i\psi_\mu(x)U_\mu(x),
\nonumber\\
   &Q\psi_\mu(x)=i\psi_\mu(x)\psi_\mu(x)
   -i\left(\phi(x)-U_\mu(x)\phi(x+a\hat\mu)U_\mu(x)^{-1}\right),
\nonumber\\
   &Q\phi(x)=0,
\nonumber\\
   &Q\bar\phi(x)=\eta(x),\qquad
   Q\eta(x)=\left[\phi(x),\bar\phi(x)\right],
\nonumber\\
   &Q\chi(x)=H(x),\qquad QH(x)=\left[\phi(x),\chi(x)\right],
\label{twoxone}
\end{align}
where $U_\mu(x)\in SU(k)$ are conventional gauge link variables, $\phi(x)$ is a
complex scalar field ($\bar\phi(x)$ is its complex conjugate),
$\Psi(x)^T\equiv(\psi_0(x),\psi_1(x),\chi(x),(1/2)\eta(x))$ are fermionic
fields and $H(x)$ is the auxiliary field; $a$ and $\hat\mu$ respectively denote
the lattice spacing and a unit vector along the direction~$\mu$ ($\mu=0$ or~1).
One confirms that the square of above transformation~(\ref{twoxone}) is a
lattice gauge transformation with the parameter~$\phi(x)$, $Q^2=\delta_\phi$;
$Q$ is thus nilpotent on gauge invariant combinations. The exact invariance of
the lattice action~$S_{\text{2DSYM}}^{\text{LAT}}$ under~$Q$ is then realized by
defining it in a $Q$-exact form,
\begin{equation}
   S_{\text{2DSYM}}^{\text{LAT}}=QX,
\label{twoxtwo}
\end{equation}
where $X$ is a certain gauge invariant combination whose explicit form can be
found in Ref.~\cite{Sugino:2004qd}.

The super transformation in the target continuum theory, 2D $\mathcal{N}=(2,2)$
SYM, has four spinor components, $(Q^{(0)},Q^{(1)},\tilde Q,Q)$, and above
transformation~(\ref{twoxone}) is a lattice transcription of the continuum
$Q$~transformation. The lattice formulation however does not possess invariance
under other three transformations, $Q^{(0)}$, $Q^{(1)}$ and~$\tilde Q$, and a
crucial issue is whether the invariance under these three transformations is
restored in the continuum limit or not.

The present lattice formulation possesses two exact bosonic symmetries.
Important in what follows is the $U(1)_A$ symmetry, under which\footnote{%
We adopt the convention
\begin{equation}
   \Gamma_0=\begin{pmatrix}
     -i\sigma_1&0\\
     0&i\sigma_1
   \end{pmatrix},\quad
   \Gamma_1=\begin{pmatrix}
     i\sigma_3&0\\
     0&-i\sigma_3
   \end{pmatrix},\quad
   \Gamma_2=\begin{pmatrix}
     0&-i\\
     -i&0
   \end{pmatrix},\quad
   \Gamma_3=C=\begin{pmatrix}
     0&1\\
     -1&0
   \end{pmatrix},
\label{twoxthree}
\end{equation}
and $\Gamma_5\equiv\Gamma_0\Gamma_1\Gamma_2\Gamma_3=
\begin{pmatrix}
     \sigma_2&0\\
     0&-\sigma_2
   \end{pmatrix}$.}%
\footnote{Another manifest bosonic symmetry is the invariance under a ``flip''
of the 0- and 1-axes~\cite{Sugino:2004qd}, under which
\begin{align}
   &U_0(x)\to U_1(\tilde x),\qquad U_1(x)\to U_0(\tilde x),\qquad
   H(x)\to-H(\tilde x),
\nonumber\\
   &\phi(x)\to\phi(\tilde x),\qquad\bar\phi(x)\to\bar\phi(\tilde x),\qquad
\nonumber\\
   &\Psi(x)\to\mathcal{F}\Psi(\tilde x),\qquad
   \mathcal{F}\equiv\frac{1}{2}(i+\Gamma_5)(\Gamma_0-\Gamma_1),
\label{twoxfour}
\end{align}
where $\tilde x\equiv(x_1,x_0)$ for $x\equiv(x_0,x_1)$. We, however, do not
employ this 0-1 flip symmetry in the present analysis.}
\begin{align}
   &\Psi(x)\to\exp\left(\alpha\Gamma_2\Gamma_3\right)\Psi(x),
\nonumber\\
   &\phi(x)\to\exp\left(2i\alpha\right)\phi(x),\qquad
   \bar\phi(x)\to\exp\left(-2i\alpha\right)\bar\phi(x).
\label{twoxfive}
\end{align}
From Eq.~(\ref{twoxone}), we see that the $Q$~transformation has the $U(1)_A$
charge $+1$, i.e., $Q\to e^{i\alpha}Q$ under $U(1)_A$. Also, the combination~$X$
in~Eq.~(\ref{twoxtwo}) has $U(1)_A$ charge~$-1$ and thus the lattice
action~$S_{\text{2DSYM}}^{\text{LAT}}$ is neutral under $U(1)_A$ as it should be
($U(1)_A$ is a manifest lattice symmetry).\footnote{Note that $U(1)_A$ is not
anomalous in 2D SYM.} Although the target continuum theory possesses other
$R$-symmetries, the $U(1)_V$ symmetry and a $Z_2$~symmetry, the present lattice
formulation is not invariant under these two.

Now, the most transparent way to examine the restoration of SUSY in the
continuum limit would be to consider a WT identity associated with SUSY. To
derive a corresponding identity in the present lattice formulation, we first
define a lattice analogue of continuum fermionic transformations other
than~$Q$, i.e., $Q^{(0)}$, $Q^{(1)}$ and~$\tilde Q$.

For this, it is convenient to introduce two bosonic transformations $R$
and~$S$: $R$ is defined by
\begin{align}
   R:&\Psi(x)\to i\Gamma_2\Psi(x),\qquad\phi(x)\to-\bar\phi(x),\qquad
   \bar\phi(x)\to-\phi(x),
\nonumber\\
   &H(x)\to-H(x)+i\hat\Phi(x),
\label{twoxsix}
\end{align}
where $\hat\Phi(x)$ is a particular combination~\cite{Sugino:2004qd} of the
plaquette variables, whose continuum limit is the 2D field strength
$2a^2F_{01}(x)$ ($a^2F_{01}(x)\equiv a\partial_0 A_1(x)-a\partial_1 A_0(x)
+i[A_0(x),A_1(x)]$). $S$ is defined by
\begin{equation}
   S:\Psi(x)\to i\Gamma_5\Psi(x).
\label{twoxseven}
\end{equation}
In the continuum limit, these $R$ and~$S$ are a part of $R$-symmetries in the
continuum target theory (the former is a $Z_2$ symmetry and the latter is the
$U(1)_V$ symmetry $\Psi(x)\to\exp(i\alpha\Gamma_5)\Psi(x)$ with the angle
$\alpha=\pi/2$). We note that $R$ flips the sign of the $U(1)_A$ charge, while
$S$ does not change the $U(1)_A$ charge. In the continuum target theory,
fermionic transformations, $Q^{(0)}$, $Q^{(1)}$ and~$\tilde Q$, are related to
the $Q$~transformation by (the continuum limit of) $R$ and~$S$, as
\begin{equation}
   Q^{(0)}=RSQS^{-1}R^{-1},\qquad Q^{(1)}=RQR^{-1},\qquad\tilde Q=SQS^{-1}.
\label{twoxeight}
\end{equation}
We can thus define $Q^{(0)}$, $Q^{(1)}$ and~$\tilde Q$ transformations on the
lattice by applying relations~(\ref{twoxeight}) to lattice
$Q$~transformation~(\ref{twoxone}). A virtue of this approach is that the
covariance under $U(1)_A$ becomes manifest. In fact,
from~Eq.~(\ref{twoxeight}), it immediately follows that
$(Q^{(0)},Q^{(1)},\tilde Q,Q)\to
(e^{-i\alpha}Q^{(0)},e^{-i\alpha}Q^{(1)},e^{i\alpha}\tilde Q,e^{i\alpha}Q)$ under
$U(1)_A$~transformation~(\ref{twoxfive}). Also, from the nilpotency of~$Q$
and~Eq.~(\ref{twoxeight}), the lattice $Q^{(0)}$, $Q^{(1)}$ and~$\tilde Q$ are
individually nilpotent on gauge invariant combinations. However, since the
lattice action is not invariant under $R$ and~$S$, $Q^{(0)}$, $Q^{(1)}$
and~$\tilde Q$ are not lattice symmetries; we note
\begin{align}
   S_{\text{2DSYM}}^{\text{LAT}}&=QX
\nonumber\\
   &=Q^{(0)}RSX+\left(1-RS\right)S_{\text{2DSYM}}^{\text{LAT}}
\nonumber\\
   &=Q^{(1)}RX+\left(1-R\right)S_{\text{2DSYM}}^{\text{LAT}}
\nonumber\\
   &=\tilde QSX
   +\left(1-S\right)S_{\text{2DSYM}}^{\text{LAT}}.
\label{twoxnine}
\end{align}
In the second line above, for example, the first term~$Q^{(0)}RSX$ vanishes
under the action of~$Q^{(0)}$ because $Q^{(0)}$ is nilpotent. However, the
second term $(1-RS)S_{\text{2DSYM}}^{\text{LAT}}$ is an $O(a)$ quantity (because
this combination vanishes in the naive continuum limit owing to
$R$-symmetries in the continuum theory) that does not necessarily vanish
under~$Q^{(0)}$. We note that each term in~Eq.~(\ref{twoxnine}), such as
$Q^{(0)}RSX$ or~$\left(1-RS\right)S_{\text{2DSYM}}^{\text{LAT}}$, is manifestly
neutral under~$U(1)_A$.

We are now ready to derive the lattice SUSY WT identity. We define a would-be
super transformation on the lattice~$\delta$ by
\begin{equation}
   \delta\equiv\frac{1}{a^{1/2}}
   \left(\varepsilon^{(0)}Q^{(0)}+\varepsilon^{(1)}Q^{(1)}
   +\tilde\varepsilon\tilde Q+\varepsilon Q\right),\qquad
   \epsilon\equiv-(\varepsilon^{(0)},\varepsilon^{(1)},
   \tilde\varepsilon,\varepsilon),
\label{twoxten}
\end{equation}
where $(\varepsilon^{(0)},\varepsilon^{(1)},\tilde\varepsilon,\varepsilon)$ are
Grassmann parameters. A WT identity can be derived as usual by employing a
localized version of~$\delta$, that is defined by $\epsilon\to\epsilon(x)$
in~Eq.~(\ref{twoxten}). We note that the identity
\begin{equation}
   \int\left[d(\text{fields})\right]\delta
   \left[e^{-S_{\text{2DSYM}}^{\text{LAT}}-S_{\text{mass}}^{\text{LAT}}}\,
   \mathcal{O}(y_1,\ldots,y_n)\right]=0,
\label{twoxeleven}
\end{equation}
holds for any multi-local operator $\mathcal{O}(y_1,\ldots,y_n)$. As
in~Ref.~\cite{Kanamori:2008bk}, here we have introduced a scalar mass term
\begin{equation}
   S_{\text{mass}}^{\text{LAT}}\equiv
   \frac{\mu^2}{g^2}\sum_x\tr\left[\bar\phi(x)\phi(x)\right],
\label{twoxtwelve}
\end{equation}
which explicitly breaks SUSY. Identity~(\ref{twoxeleven}) holds because the
functional integral measure~$[d(\text{fields})]$
(see~Ref.~\cite{Sugino:2004qd}) is invariant under the shift of integration
variables induced by the localized~$\delta$; $[d(\text{fields})]$ is obviously
invariant under $R$ and~$S$ and it is invariant also under the shift of
variables induced by~$Q$~\cite{Sugino:2006uf}.

We now set
\begin{equation}
   \delta S_{\text{2DSYM}}^{\text{LAT}}\equiv-ia^2\sum_x\epsilon(x)^T
   \left[-\partial_\mu^*s_\mu(x)+B(x)\right],
\label{twoxthirteen}
\end{equation}
where $\partial_\mu^*$ denotes the backward difference operator:
$\partial_\mu^*f(x)\equiv(1/a)(f(x)-f(x-a\hat\mu))$. $s_\mu(x)$ is
a lattice counterpart of the supercurrent and the breaking term~$B(x)$ arises
from the non-invariance of the lattice action~$S_{\text{2DSYM}}^{\text{LAT}}$
under~$\delta$. The separation of $\delta S_{\text{2DSYM}}^{\text{LAT}}$ into
$-\partial_\mu^*s_\mu(x)$ and~$B(x)$ in Eq.~(\ref{twoxthirteen}) is not unique
and we fix this ambiguity as follows: In considering terms
in~$\delta S_{\text{2DSYM}}^{\text{LAT}}$ that are proportional
to~$\varepsilon^{(0)}(x)$, for example, we use the decomposition in the second
line of~Eq.~(\ref{twoxnine}). A part of the Noether current
$-\partial_\mu^*s_\mu(x)$ is read off from the variation of the first term
$Q^{(0)}RSX$ (that is invariant under the global $Q^{(0)}$ transformation),
while the breaking effect $B(x)$ is read off from the variation of the second
term~$(1-RS)S_{\text{2DSYM}}^{\text{LAT}}$ that is $O(a)$. Similarly, for
$\varepsilon^{(1)}(x)$ (for~$\tilde\varepsilon(x)$), we use the decomposition
in the third (fourth) line of~Eq.~(\ref{twoxnine}). For~$\varepsilon(x)$, since
$S_{\text{2DSYM}}^{\text{LAT}}=QX$ is manifestly invariant under~$Q$, we can define
a conserved Noether current without the breaking term. That is, the breaking
term has the structure
\begin{equation}
   B(x)^T=(*,*,*,0).
\label{twoxfourteen}
\end{equation}

Since, for example, both $Q^{(0)}RSX$ and~$(1-RS)S_{\text{2DSYM}}^{\text{LAT}}$ are
neutral under~$U(1)_A$, and $Q^{(0)}$ has a definite $U(1)_A$ charge~$-1$, the
above prescription provides the supercurrent~$s_\mu(x)$ and the breaking term
$B(x)$ which are covariant under~$U(1)_A$. That is, we have
$s_\mu(x)\to\exp(-\alpha\Gamma_2\Gamma_3)s_\mu(x)$
and~$B(x)\to\exp(-\alpha\Gamma_2\Gamma_3)B(x)$ under $U(1)_A$.\footnote{
It turns out that the supercurrent and the breaking term are covariant also
under flip transformation~(\ref{twoxfour}) as,
$s_\mu(x)\to\mathcal{F}s_\mu(\tilde x)$ and~$B(x)\to\mathcal{F}B(\tilde x)$.}
We do not need the (quite complicated) explicit expression of $s_\mu(x)$
and~$B(x)$ in what follows. A naive continuum limit of the lattice supercurrent
reads,
\begin{align}
   s_\mu(x)&=-\frac{1}{a^{7/2}}\frac{2}{g^2}C\Bigl(
   -i\Gamma_0\Gamma_1\Gamma_\mu\tr\left[H(x)\Psi(x)\right]
\nonumber\\
   &\qquad\qquad\qquad{}
   -i\Gamma_\nu\Gamma_\uparrow\Gamma_\mu\tr\left[aD_\nu\phi(x)\Psi(x)\right]
   -i\Gamma_\nu\Gamma_\downarrow\Gamma_\mu\tr\left[aD_\nu\bar\phi(x)\Psi(x)\right]
\nonumber\\
   &\qquad\qquad\qquad{}
   -\frac{i}{2}\left[\Gamma_\uparrow,\Gamma_\downarrow\right]\Gamma_\mu
   \tr\left[\left[\phi(x),\bar\phi(x)\right]\Psi(x)\right]+O(a)\Bigr),
\label{twoxfifteen}
\end{align}
where $g$ is the 2D gauge coupling constant and
$\Gamma_{\uparrow,\downarrow}\equiv(i/2)(\Gamma_2\mp i\Gamma_3)$;
$D_\mu$~denotes the covariant derivative with respect to the adjoint
representation, $aD_\mu\equiv a\partial_\mu+i[A_\mu,\cdot]$.

For the scalar mass term, setting
\begin{equation}
   \delta S_{\text{mass}}^{\text{LAT}}
   \equiv-ia^2\sum_x\epsilon(x)^T\frac{\mu^2}{g^2}f(x),
\label{twoxsixteen}
\end{equation}
we have
\begin{equation}
   f(x)=\frac{1}{a^{5/2}}2iC\left(
   \Gamma_\uparrow\tr\left[\phi(x)\Psi(x)\right]
   +\Gamma_\downarrow\tr\left[\bar\phi(x)\Psi(x)\right]
   \right).
\label{twoxseventeen}
\end{equation}

By combining Eqs.~(\ref{twoxeleven}), (\ref{twoxthirteen})
and~(\ref{twoxsixteen}) and noting that the function~$\epsilon(x)$ is
arbitrary, we have the lattice SUSY WT identity,
\begin{align}
   &\partial_\mu^*\left\langle s_\mu(x)\mathcal{O}(y_1,\ldots,y_n)\right\rangle
\nonumber\\
   &=\frac{\mu^2}{g^2}\left\langle f(x)\mathcal{O}(y_1,\ldots,y_n)\right\rangle
   -i\frac{\delta}{\delta\epsilon(x)}
   \left\langle\mathcal{O}(y_1,\ldots,y_n)\right\rangle
   +\left\langle B(x)\mathcal{O}(y_1,\ldots,y_n)\right\rangle,
\label{twoxeighteen}
\end{align}
where $\delta\mathcal{O}(y_1,\ldots,y_n)\equiv a^2\sum_x\epsilon(x)^T
(\delta/\delta\epsilon(x))\mathcal{O}(y_1,\ldots,y_n)$.
We emphasize that this identity holds irrespective of the boundary conditions,
because we could assume that the localized parameter $\epsilon(x)$ has a
compact support which does not overlap with the boundary.

Compared with the SUSY WT identity expected in the continuum target theory,
lattice SUSY WT identity~(\ref{twoxeighteen}) has additional contribution owing
to the breaking term~$B(x)$. $B(x)$ is an $O(a)$ lattice artifact. However,
it can generally become $O(1)$ in correlation functions when combined with the
ultraviolet divergence. In the next section, by employing formal perturbation
theory, we discuss how $B(x)$ behaves in the continuum limit.

\section{Operator mixing and application of the lattice SUSY WT identity}
\label{sec:3}
In perturbation theory, one has to introduce the gauge fixing and the
associated Faddeev-Popov ghost term (see, for example,
Ref.~\cite{Luscher:1988sd}). Since these are not invariant under super
transformations, they generally give rise to additional contribution to lattice
SUSY WT identity~(\ref{twoxeighteen}). Also, if the multi-local operator
$\mathcal{O}(y_1,\ldots,y_n)$ in~Eq.~(\ref{twoxeighteen}) is not gauge
invariant (just for a collection of elementary fields), one has to take into
account the operator mixing with gauge non-invariant
operators~\cite{Taniguchi:1999fc,Farchioni:2001yr}. To avoid these
complications, in the present note, we assume that the multi-local operator
$\mathcal{O}(y_1,\ldots,y_n)$ in~Eq.~(\ref{twoxeighteen}) is a collection of
gauge invariant composite operators.\footnote{We can regard the Faddeev-Popov
ghosts and the Nakanishi-Lautrup auxiliary field as SUSY singlet. Then, since
the operations~$Q$, $R$ and~$S$ possess gauge-invariant meaning, lattice super
transformation~(\ref{twoxten}) and the lattice BRST
transformation~\cite{Luscher:1988sd} commute. This implies that SUSY variation
of the gauge fixing and the Faddeev-Popov terms is BRST exact and does not
contribute to lattice SUSY WT identity~(\ref{twoxeighteen}) if the
operator~$\mathcal{O}$ is gauge (and thus BRST) invariant.}

We first consider the case in which the point $x$ differs from $y_1$, \dots,
$y_n$ in~Eq.~(\ref{twoxeighteen}). In this case, the contact term (the second
term in the right-hand side of in~Eq.~(\ref{twoxeighteen})) is absent and, in
the continuum limit, the operator $B(x)$ may mix with gauge invariant fermionic
local operators whose mass dimension is equal to or less
than~$5/2$.\footnote{$B(x)$ has the structure that $1/g^2$ times a dimension
$9/2$ operator. Since the loop expansion parameter in the present system
is~$g^2$ and it has the mass dimension~2, in the continuum limit, $B(x)$ mixes 
with operators whose mass dimension is equal to or less than~$5/2$, as a result
of radiative corrections in 1PI diagrams containing~$B(x)$.} Taking into
account the covariance of~$B(x)$ under~$U(1)_A$~(\ref{twoxfive}),
$B(x)\to\exp(-\alpha\Gamma_2\Gamma_3)B(x)$, one sees that a possible operator
with which $B(x)$ can mix is a linear combination of the following eight
operators (we have used the fact that $\tr[\Psi(x)]\equiv0$ for the gauge
group~$SU(k)$)
\begin{align}
   &\frac{1}{a^{5/2}}C\Gamma_\uparrow\tr[\phi(x)\Psi(x)],\quad
   \frac{1}{a^{5/2}}C\Gamma_\mu\Gamma_\uparrow\tr[\phi(x)\Psi(x)],\quad
   \frac{1}{a^{5/2}}C\Gamma_5\Gamma_\uparrow\tr[\phi(x)\Psi(x)],
\nonumber\\
   &\frac{1}{a^{5/2}}C\Gamma_\downarrow\tr[\bar\phi(x)\Psi(x)],\quad
   \frac{1}{a^{5/2}}C\Gamma_\mu\Gamma_\downarrow\tr[\bar\phi(x)\Psi(x)],\quad
   \frac{1}{a^{5/2}}C\Gamma_5\Gamma_\downarrow\tr[\bar\phi(x)\Psi(x)].
\label{threexone}
\end{align}
We further \emph{assume\/} that supersymmetry itself has no intrinsic anomaly.
That is, we assume that in the continuum limit the breaking effect can be
removed by local counterterms. Then only possible mixing turns to be
$B(x)\xrightarrow{a\to0}cf(x)$, where $c$ is a constant and $f(x)$ is given
by~Eq.~(\ref{twoxseventeen}). In fact, this combination may be removed by the
super transformation of a scalar mass term. However, because of
structure~(\ref{twoxfourteen}) (that follows from the $Q$-invariance of the
formulation), the constant~$c$ must vanish. In this way, we see that
$B(x)\xrightarrow{a\to0}0$ and the continuum limit of the lattice SUSY WT
identity becomes
\begin{align}
   \partial_\mu\left\langle s_\mu(x)\mathcal{O}(y_1,\ldots,y_n)\right\rangle
   =\frac{\mu^2}{g^2}\left\langle f(x)\mathcal{O}(y_1,\ldots,y_n)\right\rangle,
\label{threextwo}
\end{align}
when the point $x$ differs from $y_1$, \dots, $y_n$. This relation shows that
the lattice supercurrent $s_\mu(x)$, without any renormalization, reproduces in
the continuum limit a relation expected in the target continuum
theory.\footnote{To show this, we thus used the $Q$ and $U(1)_A$ symmetries of
the lattice formulation \emph{and\/} the absence of an intrinsic SUSY anomaly
in the target theory. It might appear that we needed a further assumption on
the absence of SUSY anomaly compared with the argument
in~Ref.~\cite{Sugino:2003yb}. However, one should note that this assumption is
implicitly made also in~Ref.~\cite{Sugino:2003yb}. Actually,
in~Ref.~\cite{Sugino:2003yb}, the possibility of SUSY breaking arising from
\emph{non-local\/} terms is not taken into account from the beginning. If one
does not like to accept a priori the absence of SUSY anomaly in this system, it
would be possible to confirm this by explicit (one-loop) perturbative
consideration.}
Such a supercurrent on the lattice is however not unique. In fact, let
$s_\mu'(x)$ be an appropriately-chosen another lattice supercurrent such that
$\Delta s_\mu(x)\equiv s_\mu'(x)-s_\mu(x)=O(a)$ is gauge invariant. Then
$\Delta s_\mu(x)$ can mix with gauge invariant dimension~$3/2$ fermionic local
operators. Only possible operator mixing is thus
$\Delta s_\mu(x)\xrightarrow{a\to0}M\tr[\Psi(x)]\equiv0$ ($M$ being a certain
$4\times4$ matrix) for the gauge group $SU(k)$. This shows that a precise
choice of a lattice supercurrent is not relevant for identity~(\ref{threextwo})
to hold in the continuum limit.

This corresponds precisely to the situation studied
in~Ref.~\cite{Kanamori:2008bk}. There, the authors employed an
appropriately-chosen lattice supercurrent $s_\mu'(x)$ that is different
from~$s_\mu(x)$ by an $O(a)$ amount. The composite operator was
\begin{equation}
   \mathcal{O}(y)=f_\nu(y)\equiv-\frac{1}{2g^2}\Gamma_\nu C^{-1}f(y),
\label{threexthree}
\end{equation}
and the restoration of relation~(\ref{threextwo}) with $x\neq y$ in the
continuum limit was observed by means of the Monte Carlo simulation. This
demonstrated the SUSY restoration in a nonperturbative level.

Usually, from a WT identity such as~(\ref{threextwo}) that does not contain the
contact term, i.e., the second term of the right-hand side
of~Eq.~(\ref{twoxeighteen}), one cannot conclude that the current
operator~$s_\mu(x)$ is finite or correctly-normalized. In our present 2D case,
fortunately, we can directly see that the supercurrent~$s_\mu(x)$ and the
operator~$f(x)$ are finite operators which do not require nontrivial
renormalization. One can readily see that 1PI diagrams that contain $s_\mu(x)$
or~$f(x)$ are ultraviolet finite except one-loop diagrams being proportional
to~$\tr[\Psi(x)]\equiv0$. Thus, the above supercurrent, $s_\mu(x)$
or~$s_\mu'(x)$, is a correctly-normalized, finite operator.

As an interesting application of lattice SUSY WT identity~(\ref{twoxeighteen})
is obtained by taking an appropriately-chosen lattice supercurrent $s_\mu'(y)$
itself as the composite operator:
\begin{equation}
   \mathcal{O}(y)=\left(s_0'\right)_{i=1}(y),
\label{threexfour}
\end{equation}
where $i$ refers to the spinor index. The $i=1$ component of the supercurrent
corresponds to a Noether current associated with the fermionic
transformation~$Q^{(0)}$. Then, assuming that a naive $\mu^2\to0$ limit can be
taken in lattice SUSY WT identity~(\ref{twoxeighteen}), we have
\begin{equation}
   \partial_\mu^*\left\langle\left(s_\mu\right)_{i=4}(x)
   \left(s_0'\right)_{i=1}(y)\right\rangle
   =i\frac{1}{a^2}\delta_{x,y}
   \left\langle Q\left(s_0'\right)_{i=1}(x)\right\rangle.
\label{threexfive}
\end{equation}
Note that we have focused especially on the $i=4$ spinor component of the
lattice supercurrent~$s_\mu(x)$. Since the $i=4$ component corresponds to the
$Q$~transformation, we do not have the breaking term~$B(x)$
in~Eq.~(\ref{threexfive}) even with finite lattice spacings (recall
Eq.~(\ref{twoxfourteen})). Now, in the target continuum theory
\emph{in classical level}, the $Q$~transformation of the time component of the
Noether current associated with the $Q^{(0)}$~transformation is the hamiltonian
density, $Q\left(s_0'\right)_{i=1}(x)=2\mathcal{H}(x)$, as is consistent with
the SUSY algebra, $\{Q,Q^{(0)}\}=-2i\partial_0+2\delta_{A_0}$. Therefore, it is
quite natural to regard the right-hand side of~Eq.~(\ref{threexfive}) as the
expectation value of the hamiltonian density \emph{in quantum theory}:
\begin{equation}
   \left\langle Q\left(s_0'\right)_{i=1}(x)\right\rangle\equiv
   2\left\langle\mathcal{H}(x)\right\rangle.
\label{threexsix}
\end{equation}
This is precisely the prescription advocated
in~Refs.~\cite{Kanamori:2007ye,Kanamori:2007yx} for the
hamiltonian density in the present lattice formulation. The reasoning for this
prescription in~Refs.~\cite{Kanamori:2007ye,Kanamori:2007yx} was based on a
topological property of the Witten index. Here, we arrived at the identical
prescription from an argument of the operator algebra among
correctly-normalized supercurrents. This provides another justification for the
prescription in~Refs.~\cite{Kanamori:2007ye,Kanamori:2007yx}.

One might wonder to what extent the definition of the hamiltonian
density~$\mathcal{H}(x)$ in Eq.~(\ref{threexsix}) is affected by a choice
of the supercurrent~$s_0'(y)$ in~Eqs.~(\ref{threexfive}) and~(\ref{threexsix}).
Let $\Delta s_\nu'(y)\equiv s_\nu''(y)-s_\nu'(y)=O(a)$, where $s_\nu''(y)$
denotes a yet another (gauge invariant) lattice supercurrent. An argument
similar to above then shows that this does not contribute to the left-hand side
of~Eq.~(\ref{threexfive}), 
$\Delta s_0'(y)\xrightarrow{a\to0}0$ when~$x\neq y$. $\Delta s_0'(y)$ can
contribute only when the positions of two composite operators coincide,
i.e., when~$x=y$. From a dimensional analysis, a possible effect of the
difference in the left-hand side of~(\ref{threexfive}) is thus
$\partial_\mu^*\langle(s_\mu)_{i=4}(x)(\Delta s_0')_{i=1}(y)\rangle
\xrightarrow{a\to0}
(d_{00}(\partial_0)^2+d_{01}\partial_0\partial_1+d_{11}(\partial_1)^2)
\delta^2(x-y)$, where $d_{\alpha\beta}$ are constants. However, since the
continuum limit of the difference in the right-hand side
of~Eq.~(\ref{threexfive}) is proportional to~$\delta^2(x-y)$ without
derivative, we conclude that $d_{00}=d_{01}=d_{11}=0$; the continuum limit of
the hamiltonian density is not affected by a choice of~$s_0'(y)$.

On the basis of this prescription for the hamiltonian density,
in~Refs.~\cite{Kanamori:2007ye,Kanamori:2007yx} and more extensively
in~Ref.~\cite{Kanamori:2009dk}, the vacuum energy density of 2D
$\mathcal{N}=(2,2)$ SYM has been numerically computed. This would provide a
possible clue for a conjectured spontaneous SUSY breaking in this
system~\cite{Hori:2006dk}. Note that Eqs.~(\ref{threexfive})
and~(\ref{threexsix}) show that $\langle\mathcal{H}(x)\rangle$ is precisely the
order parameter of the SUSY breaking, in the sense that its non-zero (positive)
value ensures the massless Nambu-Goldstone fermion in the channel of the
left-hand side of~Eq.~(\ref{threexfive}).

\section{Confirmation of a SUSY WT identity in small volume lattices}
\label{sec:4}
Our discussion on the operator mixing in the previous section is somewhat
formal because perturbation theory in 2D gauge theory suffers from the infrared
divergence. For generic quantities, one cannot trust perturbation theory in
infinite volume, even if the dimensionless loop expansion parameter $(ag)^2$
becomes very small in the continuum limit.\footnote{For example, the
expectation value of the action density $\mathcal{L}$ in 2D $SU(2)$ Yang-Mills
theory, defined by the plaquette action, is given by
$\langle\mathcal{L}\rangle=(3/2)(1/a^2)-(3/32)g^2$ in the continuum limit; this
is an \emph{exact\/} expression obtained by the character expansion. On the
other hand, perturbation theory in infinite volume (see, for example,
Ref.~\cite{Heller:1984hx}) yields
$\langle\mathcal{L}\rangle=(3/2)(1/a^2)+(1/32)g^2$ to the first nontrivial
order and this is wrong. There is no real paradox here, because higher-order
perturbative corrections are infrared diverging and perturbation theory in
infinite volume itself is meaningless for this quantity.}
The infrared divergence can be avoided by putting the system into a finite box
of size~$L$ (we set the one-dimensional number of lattice points
$N\equiv L/a$) that introduces a physical energy scale to the problem. Then
perturbation theory turns out to be an asymptotic expansion with respect to
$(Lg)^2$, rather than $(ag)^2$ (the infrared divergence is reproduced as a
divergence in~$L\to\infty$). Therefore, we may always employ perturbation
theory, if volume of the system is small enough measured in the gauge coupling.
Certainly, perturbation theory cannot completely substitute Monte Carlo
simulations, if one is interested in low-energy physics in large physical
volume.

In perturbation theory in a finite box, however, another complication arises;
depending on the boundary condition, constant modes of various (perturbatively)
massless fields may survive. One cannot apply the conventional perturbation
theory to those constant modes because they do not have a quadratic kinetic
term; they are rather subject of nonperturbative integrations. In the context
of a lattice formulation of 2D $\mathcal{N}=(2,2)$ SYM
of~Ref.~\cite{Cohen:2003xe}, the two-point correlation function of scalar
fields at zero momentum has been studied by combining one-loop perturbation
theory and nonperturbative integrations over constant
modes~\cite{Onogi:2005cz}. (For the nonperturbative integration, the technique
in~Ref.~\cite{Suyama:1997ig} was employed.) In what follows, we confirm a SUSY
WT identity examined in~Ref.~\cite{Kanamori:2008bk} by using this
``semi-perturbative'' treatment to the first nontrivial order. This analytical
study supplements the formal argument in the previous section. Compared with
the Monte Carlo study~\cite{Kanamori:2008bk}, this analytical study is
advantageous in that it is free from statistical/systematic errors. We consider
the case in which fermionic fields obey the periodic boundary condition along
the temporal direction; for this case no definite conclusion was obtained
in~Ref.~\cite{Kanamori:2008bk} owing to large statistical errors.

We thus first parametrize the link variables by gauge potentials as
$U_\mu(x)=\exp(iA_\mu(x))$. We introduce the measure term~\cite{Kawai:1980ja}
and the gauge fixing and the Faddeev-Popov ghost terms~\cite{Luscher:1988sd}.
We then decompose lattice fields as\footnote{For simplicity of calculation, we
assumed that $N$ is an odd integer.}
\begin{equation}
   A_\mu(x)=\sum_ke^{ikx/a}\,\tilde A_\mu(k),\qquad
   k_\mu\equiv\frac{2\pi n_\mu}{N},\quad
   n_\mu=0,1,2,\dots,N-1,
\label{fourxone}
\end{equation}
and similar expressions for other fields. For modes with $k_\mu\neq0$, we can
apply the perturbative expansion. For constant modes with which $k_\mu=0$, a
perturbative expansion is impossible and one has to generally carry out the
integration in a nonperturbative way. It can be seen from the lattice action,
the expectation value of $\tilde A_\mu(0)$ and $\tilde\phi(0)$ is
$O((ag)^{1/2})$ while the expectation value of $\tilde\Psi(0)$ (that is present
for the periodic boundary condition) is $O((ag)^{3/4})$.

Now, we are interested in whether a SUSY WT identity of the form
of~Eq.~(\ref{threextwo})~\cite{Kanamori:2008bk}
\begin{equation}
   \partial_\mu\left\langle s_\mu(x)f_\nu(y)\right\rangle
   =\frac{\mu^2}{g^2}\left\langle f(x)f_\nu(y)\right\rangle,
   \qquad\text{for $x\neq y$,}
\label{fourxtwo}
\end{equation}
where the operators $s_\mu(x)$, $f_\nu(y)$ and~$f(x)$ are given
by~Eqs.~(\ref{twoxfifteen}), (\ref{threexthree}) and~(\ref{twoxseventeen}),
respectively, holds in the continuum limit or not. We thus decompose
composite operators in the left-hand side
$\left\langle s_\mu(x)f_\nu(y)\right\rangle$ into constant modes and
non-constant modes. We neglect ultraviolet finite diagrams because these should
not modify the identity in the continuum limit.\footnote{Note that the
integrations over constant modes do not produce the ultraviolet divergence.}
Then taking into account the order-counting elucidated above, it turns out that
the lowest nontrivial order contribution to this function is~$O((ag)^{3/2})$. It
is given by: Fermion fields $\Psi(x)$ and~$\Psi(y)$ in composite operators are
replaced by the constant mode~$\tilde\Psi(0)$ and scalar fields in composite
operators are connected by the scalar two-point function with one-loop
self-energy corrections. By applying $\partial_\mu$ to this lowest-order term,
one finds
\begin{equation}
   \partial_\mu\left\langle s_\mu(x)f_\nu(y)\right\rangle
   =\left(\frac{\mu^2}{g^2}+\mathcal{C}\right)
   \left\langle f(x)f_\nu(y)\right\rangle,
   \qquad\text{for $x\neq y$,}
\label{fourxthree}
\end{equation}
to $O((ag)^{3/2})$, where the constant $\mathcal{C}$ is given by the one-loop
self-energy of scalar fields arising from integrations over non-constant modes.
Although the self-energy itself depends on the external momentum, the
dependence is higher order in $(ag)^2$ for a dimensional reason; we can thus
set the external momentum zero and regard the self-energy as a constant. In the
function $\langle f(x)f_\nu(y)\rangle$ in the right-hand side
of~Eq.~(\ref{fourxthree}), fermion fields $\Psi(x)$ and~$\Psi(y)$ in composite
operators are also replaced by the constant mode~$\tilde\Psi(0)$ and scalar
fields in composite operators are connected by the scalar two-point function to
the one-loop order.

Eq.~(\ref{fourxthree}) shows that if $\mathcal{C}\neq0$ in the continuum limit
then the expected SUSY WT identity is not restored. A straightforward one-loop
calculation yields
\begin{equation}
   \mathcal{C}=k\frac{2}{N^2}
   \sum_{(n_0,n_1)\neq(0,0)}\left[
   \frac{1}{2}\left(1+\frac{1}{\lambda}\right)\frac{1}{\hat k^2}
   +\frac{1}{2}\left(1-\frac{1}{\lambda}\right)\frac{1}{\hat k^2+a^2\mu^2}
   -\frac{1}{\hat k^2}
   \right],
\label{fourxfour}
\end{equation}
where $\lambda$ denotes the gauge parameter, $\mu^2$ is the scalar
mass-squared, $\hat k^2\equiv\sum_{\mu=0}^1(\hat k_\mu)^2$
and~$\hat k_\mu\equiv2\sin(k_\mu/2)$. In the square brackets
of~Eq.~(\ref{fourxfour}), the first term is the contribution of the gauge loop,
the second is the scalar-gauge loop and the third is the fermions'
contribution. In the second term, we can neglect $a^2\mu^2=(\mu^2/g^2)(ag)^2$
because this is higher order in~$(ag)^2$. In this way, we
have~$\mathcal{C}=0$.\footnote{
Although it is not relevant to Eq.~(\ref{fourxtwo}) in the lowest order, the
one-loop self-energy of the gauge field is also of interest because power
counting tells that it is also ultraviolet diverging. Writing the one-loop
effective action of~$\tilde A(0)$ as
$S_{\text{eff}}=\sum_{\mu,\nu=0}^1\mathcal{C}_{\mu\nu}N^2
\tr[\tilde A_\mu(0)\tilde A_\nu(0)]$, a somewhat lengthy calculation shows
\begin{equation}
   \mathcal{C}_{\mu\nu}
   =k\delta_{\mu\nu}\frac{1}{N^2}
   \sum_{(n_0,n_1)\neq(0,0)}\sum_{\rho=0}^1\frac{\partial}{\partial k_\rho}
   \left[
   0+\frac{\ring k_\rho}{\hat k^2+a^2\mu^2}-\frac{\ring k_\rho}{\hat k^2}
   \right],
\end{equation}
where $\ring k_\rho\equiv\sin k_\rho$. Thus $\mathcal{C}_{\mu\nu}=0$, if
$a^2\mu^2$ in the second term is neglected as a higher order correction. It is
interesting to note that, despite the underlying gauge invariance, this
expression itself could not vanish for finite $N$, if the field content was
not supersymmetric.} Combined with Eq.~(\ref{fourxthree}), this demonstrates
expected identity~(\ref{fourxtwo}) with the periodic boundary condition
to~$O((ag)^{3/2})$.

\ack

We would like to thank Yasuyuki Hatsuda, Yoshio Kikukawa, Martin L\"uscher,
Tetsuya Onogi, Fumihiko Sugino and Asato Tsuchiya for helpful discussions.
Discussions during the YITP workshop, ``Development of Quantum Field Theory and
String Theory'' (YITP-W-09-04), were very useful to complete this work and we
would like to thank the Yukawa Institute for Theoretical Physics at Kyoto
University for the hospitality.
The work of H.S.\ is supported in part by a Grant-in-Aid for Scientific
Research, 18540305.



\end{document}